\documentstyle[preprint,pre,aps,eqsecnum]{revtex}

\def\spose#1{\hbox to 0pt{#1\hss}}
\def\ltapprox{\mathrel{\spose{\lower 3pt\hbox{$\mathchar"218$}}
 \raise 2.0pt\hbox{$\mathchar"13C$}}}
\def\gtapprox{\mathrel{\spose{\lower 3pt\hbox{$\mathchar"218$}}
 \raise 2.0pt\hbox{$\mathchar"13E$}}}

\begin{document}

\draft
\preprint{IFUP-TH 69/95}
\draft
\title{Testing the heating method with perturbation theory}
\author{B. All\'es, M. Beccaria and F. Farchioni}
\address{
Dipartimento di Fisica dell'Universit\`a  and INFN\\
Piazza Torricelli 2, I-56100 Pisa, Italy\\
}
\maketitle
\begin{abstract}
The renormalization constants present 
in the lattice evaluation of the topological susceptibility
can be non-perturbatively calculated by using the so-called
heating method. We test this method 
for the $O(3)$ non-linear $\sigma$-model in two dimensions.
We work in a regime where perturbative calculations are exact
and useful to check the values obtained from the heating method.
The result of the test is positive and it clarifies 
some features concerning the method.
Our procedure also allows a rather accurate determination 
of the first perturbative coefficients.
\end{abstract}

\vskip 5mm


\section{Introduction}

\vskip 5mm

Matrix elements of local operators can be determined by lattice
techniques. The corresponding Monte Carlo simulation provides the 
continuum value of the
matrix element modified by lattice finite renormalizations.
The renormalization constants are usually evaluated perturbatively.
They can also be evaluated by using non-perturbative methods.
These methods yield the 
renormalization constants at any value of the bare coupling $g_0$ 
avoiding the uncertainties derived from the knowledge of only the few
first terms in the perturbative series and from 
the asymptotic character of the series.

A well-known example of
calculation of renormalization constants happens during
the determination of the
topological susceptibility in QCD or in ${\rm CP}^{N-1}$ models. The 
Monte Carlo signal for this quantity is \cite{chinpb,chiprd}
\begin{equation}
\chi^{\rm latt} = a^d(g_0) Z(g_0)^2 \chi + a^d(g_0) 
A(g_0){\left\langle T \right\rangle}_{NP} + P(g_0).
\end{equation}
In this equation $d$ is the space-time dimension, $a$ is 
the lattice spacing (related to the bare coupling
through the beta function), ${\left\langle T \right\rangle}_{NP}$ is the
non-perturbative part of the vacuum expectation value of the trace of
the energy-momentum tensor and
$Z(g_0)$, $A(g_0)$ and $P(g_0)$ are finite renormalizations. The topological
susceptibility on the lattice is defined as
\begin{equation}
\chi^{\rm latt} = \frac{1}{V} \langle (Q^{\rm latt})^2 \rangle,
\end{equation}
where $Q^{\rm latt}$ is a suitable 
definition of the topological charge operator on
the lattice and $V$ is the space-time volume.

The calculation of the constants $Z(g_0)$, $A(g_0)$ and $P(g_0)$ can be
performed either by applying perturbation theory or by using the so-called 
heating method \cite{vicari}. 
This is a direct non-perturbative method to evaluate the 
renormalization constants as it does not rely on any expansion.
It consists in heating a classical 
initial configuration having a known topological charge $Q_0$. The 
previous renormalization constants will show up during the first steps 
of thermalization because only short wave fluctuations are generated
at this stage. This method has been used in the $O(3)$ non-linear 
$\sigma$-model \cite{chiprd,plb} and in QCD \cite{alles}.

The aim of the present 
work is to check the validity of this non-perturbative
method by comparing it with perturbation theory at large $\beta$
values. 
We performed the check on the 
$O(3)$ non-linear $\sigma$-model in two dimensions.
In the continuum this model is defined by the action
\begin{equation}
S = {1 \over {2 g_0}} \int \hbox{d}^2x 
(\partial_{\mu} {\vec \phi}(x))^2,
\end{equation}
with the constraint ${\vec \phi}(x)^2=1$ for all $x$. On the lattice we
will make use of the Symanzik tree-level improved action \cite{sym}
\begin{equation}
S^{\rm latt} = - \beta \sum_{x,\mu} \left( \frac{4}{3} {\vec \phi}(x)
\cdot {\vec \phi}(x+\mu) - \frac{1}{12} {\vec \phi}(x) 
\cdot {\vec \phi}(x+2\mu) \right).
\end{equation}
Henceforth $\beta=1/g_0$. The lattice definition we used for 
the topological charge density is 
\begin{equation}
Q^{\rm latt} (x)= \frac{1}{32\pi}\epsilon_{\mu\nu}\epsilon_{ijk}
\phi^i(x) \left[ \phi^j(x+\mu) - \phi^j(x-\mu) \right]
\left[ \phi^k(x+\nu) - \phi^k(x-\nu) \right].
\end{equation}
The first two perturbative coefficients of $Z$ and $P$ for the previous
lattice action and charge density are known \cite{chiprd,plb}. 
We will test
the heating method by measuring these two renormalization constants
at very large $\beta$.

At fixed lattice size $L$ and large $\beta$ all spins tend 
to be parallel and small fluctuations around the trivial
vacuum is the only physics present in the system. The small value
of the ratio $L/\xi(\beta)$ ($\xi$ is the correlation length)
prevents disorder to appear. Therefore long wavelength fluctuations 
can hardly be generated. 
In these conditions perturbation theory becomes exact \cite{lebowitz}.
On the other hand, within our statistical errors, 
the perturbative series can be well
approximated by the first two coefficients 
if using $\beta = 100 \div 1000$.
Hence this approximate result from the perturbative series 
should reproduce the obtained value from the heating method.

We chose the $O(3)$ non-linear $\sigma$-model because it is known that
the instanton size distribution in this model favours small
instantons \cite{1ro}. Therefore an instanton heated at $\beta=1000$ 
can be well accommodated on a 
lattice size $L \sim 1000a$ which is the biggest lattice we will use.

As a byproduct we show how to use the heating method to compute the
first coefficients of the perturbative expansions in equation (1.1)
with rather high precision.

In section 2 we will review the heating method and the results and 
conclusions will be shown in sections 3 and 4.

\vskip 1cm

\section{The heating method}

\vskip 5mm

The heating method \cite{vicari,tep}
is a procedure to non-perturbatively determine the
renormalization constants in equation (1.1). We start from a given
classical configuration ${\cal C}_0$ having some known topological
charge $Q_0$. Then we construct ensembles of configurations 
$\{ {\cal C}_n \}$ obtained after performing $n$ thermalization steps on
the initial ${\cal C}_0$ configuration at some value of $\beta$.
For small $n$ it is expected that the
updating sweeps create only small fluctuations up to distances of a
few lattice spacings. 
For this purpose it is important to use a slow updating algorithm.
Convenient algorithms are Metropolis and heat-bath.
We used a heat-bath algorithm. 
If the correlation length satisfies
$\xi(\beta) \gg a$ then we can assume that the configuration contains
small statistical fluctuations on a background of topological charge
$Q_0$. The main assumption of the method is that these
fluctuations are responsible for the
renormalizations~\cite{tep}. 
Therefore if $Q_0=1$ and at each $n$ we measure 
$Q^{\rm latt}$ then after few updating steps the measured value of 
$Q$ divided by $Q_0$ will give us $Z(\beta)$. If
we measure $\chi^{\rm latt}$ and $Q_0=0$ then after few steps the
Monte Carlo signal will reach the value of $P(\beta)$.

To create the initial configuration ${\cal C}_0$ with topological
charge $Q_0=0$ we put all spins parallel to some
axis. When the initial configuration must contain a charge $Q_0=1$
then we put by hand a charge-one instanton field on the lattice. For
this purpose we use the well-known expression in the continuum for
such one-instanton field \cite{poly},
\begin{equation}
w = \frac{z - z_0}{\rho \exp(i\theta)}, \; \; \; \; \; \; \;
w \equiv \frac{\phi^1 + i \phi^3}{1 - \phi^2},
\end{equation}
where $z=x_1+i x_2$ is the coordinate on the two-dimensional euclidean
space-time and $z_0$ is the center of the instanton; we always put this 
center at the geometrical center of the lattice, $z_0=(1 + i)L/2$.
In this equation $\theta$ and $\rho$ are the orientation and size
of the instanton respectively.
We chose $\theta=\pi/4$. In the next section 
we will discuss the value for $\rho$. 
Once the instanton has been put on the lattice, we apply a relaxation
process on the configuration to settle it. Indeed the initial
charge is usually less than~1 and after the relaxation it approaches~1.
The relaxation process is repeated until the value of $Q_0$
stabilizes. 
The relaxation algorithm used was the cooling \cite{cool,chinpb}. 
We performed 30 cooling
steps.

However, even after the cooling, the background topological charge
$Q_0$ was never exactly~1. It was close but less than~1. This fact 
is not
surprising as there are no known exact instanton solutions on the lattice. 
Moreover it is known that a single 
topological charge-one field is not allowed
on a continuum torus \cite{rr} 
and we use a periodic lattice to perform our
numerical simulations. 
Therefore it is not clear whether the heating method to compute $Z$
will work or not. In particular when we normalize the measured
$Q$ with $Q_0$, $Z=Q/Q_0$, we could use either 
the lattice non-integer value of $Q_0$ or the corresponding integer
continuum value. 
Using exact toroidal biinstanton solutions \cite{rr} does
not ameliorate the situation because on the lattice
$Q_0$ is again close but less than~2.
In this work we will try to clarify these points.

The signal of our observables measured during a numerical simulation 
is less noisy at large values of the ratio $\xi/L$.
In this case, we can get rather accurate results with low statistics.
In particular, working at large values of $\beta$ allows us to observe 
the lattice size dependence of
the renormalization constants. We expect that the perturbative
tail~$P$ have a clear dependence on $L$ while the multiplicative
renormalization might be independent of $L$ as the perturbative
procedure to obtain $Z$ predicts.

\vskip 1cm

\section{Simulation and results}

\vskip 5mm

In this section we will describe the simulations performed to
determine both the multiplicative renormalization $Z$ and the
perturbative tail $P$ in equation (1.1).

At large values of $\beta$ the signal from the mixing with
the trace of the energy-momentum tensor 
(see equation 1.1) is totally negligible. Therefore we
cannot check the perturbative expansion of $A$ in equation (1.1).

\subsection{The multiplicative renormalization}

We performed several measures of the multiplicative renormalization
$Z(\beta)$. We used two lattices: $240^2$ with $\beta=100$ and
$1200^2$ with $\beta=1000$. 
At these values of $\beta$ the ratio $\xi/L$ is 
${\cal O}(10^{100 \div 1000})$.
In both cases we studied the dependence of
the result on the instanton size $\rho$. The simulations showed a
strong dependence on this parameter. For large instanton sizes 
$\rho/L \gtapprox 0.15$ the data raise
while for small sizes $\rho \ltapprox 8a$ the curve falls off
(see Figure~1). In the first case finite lattice size effects 
distort the instanton distribution cutting it off 
at the boundary of the lattice. We understand that the Monte Carlo
signal at these values of $\rho$ has no physical meaning.
In the second case the instanton is too small and after few heating
steps it 
gets dissolved in the statistical fluctuations around it. In
consequence the topological content gets lost.

In between we see a window of instanton sizes for which the Monte
Carlo signal of the heating method displays a long and 
clear plateau (see Figure~1). The value of $Z(\beta)$ is the height
of the plateau normalized to the initial charge $Q_0$.
The drift downwards 
of the curve for small $\rho$ makes it difficult to
determine the value of $Z$. This fact is reflected in the larger
error bars for small $\rho$. For large $\rho$ we chose the minimum of
the curve as the value for $Z$.
In Figure 2 and 3 we show the value of
the plateau as a function of $\rho/L$ for the two lattices we used.

For values of $\beta$ in the scaling window of the model 
($\beta \sim 1.5\div2.5$ in usual simulations) a slow drop of the
curves is seen. This is due to the creation of small size instantons
with opposite charge than the initial one \cite{inter}. 
This is a systematic error
which one has to face when applying the heating method to the $O(3)$
non-linear $\sigma$-model \cite{fpapa}. 
At large values of $\beta$, the strong 
critical slowing down prevents the creation of such instanton sea
around the background. Hence we think that the behaviour of the data
at small $\rho$ is well explained by the loss of the background
instanton in the middle of the fluctuations. 

Before extracting the value of $Z(\beta)$ from Figure~2 and~3 we
will discuss the statistics used in each run. For the runs on a
$240^2$ lattice at $\beta=100$ we performed 1000 trajectories
of 100 heating steps. The data shown in Figure~1 are the average of
these 1000 trajectories. For the $1200^2$ lattice at $\beta=1000$ we
performed 20 trajectories of 150 heating steps. In each case an
autocorrelation analysis was done. The data are correlated at
distances of $\sim 5$ heating steps.
We calculated the height and error
of the plateau taking into account this effect.
The step where the plateau starts can be determined 
by just having a look at the curve (see Figure~1).
>From this first point (say $n_0$) 
we averaged all heating steps until some $\bar n$,
$n_0 < {\bar n} \leq 100$. The height of the plateau was obtained by
looking for a stable result of this average while varying $\bar n$
from $n_0+1$ to~100.

The results of the calculation of $Z(\beta)$ for each $\rho$ are shown
in Tables I and II corresponding to Figures~2 and~3 respectively.
Getting the value of $Z(\beta=1000)$ from Figure~3 is easy.
The window of stable instanton sizes is apparent. As the lattice size
diminishes, this window shrinks and becomes a flex point of the curve.
To determine the flex point of the curve in Figure~2 we first 
interpolated the Monte Carlo points with an odd degree polynomial
and then we evaluated analytically the flex point. Increasing the
degree of the polynomial, the result for the flex stabilizes.
In Table III we show the results of this calculation as a function of
the degree of the interpolating polynomial.

The values for $Z(\beta)$ obtained from these figures are
$Z(\beta=100)=0.993159(2)$ and 
$Z(\beta=1000)=0.999317(1)$. On the other hand the
one-loop and two-loop order coefficients of this multiplicative
renormalization are 
\begin{equation}
Z(\beta)=1+\frac{z_1}{\beta} + \frac{z_2}{\beta^2} + 
{\cal O}(\frac{1}{\beta^3}) \;\;\;\;\;\;\; z_1=-0.684040,
\;\;\;\;\;\; z_2=-0.0598.
\end{equation}
The values for $z_1$ and $z_2$ were first calculated 
on finite size lattices by substituting the corresponding loop 
integrals for sums. Then these results
were extrapolated to 
infinite lattice size.
The extrapolating function was $z_i(L)=z_i+\alpha/L^m$.
This extrapolation was stable within ten digits for $m=2$ without
including logarithms of $L$ in the previous fuction.
Therefore the analytical prediction for the multiplicative
renormalization is $Z^{\rm 2-loop}(\beta=100)=0.993154$ and
$Z^{\rm 2-loop}(\beta=1000)=0.999316$. We see an excellent agreement
between the theoretical and Monte Carlo values. 

The lattice size independence of $Z$ is hard to reveal from our
method. To see a clear flex point in Figure~2 we must work on lattice
sizes satisfying $8a \ltapprox 0.15 L$ which means $L \gtapprox 60a$.
At $L=60a$ the difference 
$\mid z_1 - z_1(L=60a) \mid \sim 10^{-4}$ cannot
be seen at the values of $\beta$ we work.

All the data shown for $Z$ were equal to the Monte Carlo signal for $Q$
divided by the initial charge on
the lattice after 30 cooling steps, $Q_0$. 
We could also divide $Q$ by the corresponding continuum initial
charge, {\it i.e.}: the integer closest to the lattice value of $Q_0$.
The lattice values of $Q_0$ are $Q_0=0.99901$ on the $240^2$ lattice
at $\rho/L=0.11$ and $Q_0=0.99981$ on the $1200^2$ lattice at 
$\rho/L=0.05$. Notice that these values cannot depend on $\beta$.
The corresponding continuum values would clearly be $Q_0=1$.
Had we used these continuum values for normalizing the Monte Carlo
signal we would have got wrong results for $Z$. Indeed for the 
$240^2$ lattice at $\beta=100$ we would have obtained 0.992176(2)
while 
for the $1200^2$ lattice at $\beta=1000$ the result would have been
0.999127(1). We conclude that the normalization has to be done by
consistently using the lattice value of $Q_0$.

We can also look at the problem 
the other way round and try to determine the
value of $z_1$ from the Monte Carlo data.
Equating the Monte Carlo value for $Z(\beta=1000)$ to 
$1+z_1/1000$ we get $z_1=-0.683(1)$. We can do the same for
$Z(\beta=100)$ obtaining $z_1=-0.6841(2)$.
These results are in agreement with equation (3.1).

\subsection{The perturbative tail}

The determination of the perturbative tail $P$ 
can be performed by heating a
trivial configuration, {\it i.e.}: all spins parallel to some
previously chosen direction ${\vec v}$ in the $O(3)$ space. 
We started every trajectory with a different
direction
${\vec v}$ chosen at random.
We used rather small lattices in order to
check the lattice size dependence of $P$: $L=9a$ and $L=48a$.

The dashed line and circles 
in Figure~4 display the perturbative prediction and 
the result of the simulation on a $9^2$ lattice
at $\beta=100$. We performed 60 heating steps and $10^4$
trajectories. The perturbative tail has the form
\begin{equation}
P(\beta)=\frac{p_4}{\beta^4}+
\frac{p_5}{\beta^5}+{\cal O}(\frac{1}{\beta^6}).
\end{equation}
At $L=9a$ the values of the coefficients are 
$p_4=6.036 \times 10^{-5}$ and
$p_5=5.159 \times 10^{-5}$. To calculate the second coefficient we must
include the zero mode contribution \cite{hasen} 
which amounts to $\sim 7\%$ of the
whole term. Hence $P^{\rm 4-loop}(\beta=100,L=9a)=6.088 \times 10^{-13}$.
In Figure~4 we see an impressive agreement with the Monte Carlo value
$P(\beta=100,L=9a)=6.09(3) \times 10^{-13}$. In doing the fit to 
an horizontal line we eliminated the autocorrelations 
of the data.

The solid line and triangles in 
Figure~4 display the same perturbative prediction and simulation 
for a $48^2$ lattice. Each
trajectory consisted of 100 heating steps. For this lattice size the
perturbative coefficients are $p_4=6.804 \times 10^{-5}$ and 
$p_5=5.722 \times 10^{-5}$. Here the zero mode term is negligible.
Therefore $P^{\rm 4-loop}(\beta=100,L=48a)=6.861 \times 10^{-13}$.
The Monte Carlo result is $P(\beta=100,L=48a)=6.85(2) \times 10^{-13}$.
The agreement is again satisfactory.
We can see a lattice size dependence in the perturbative tail which
agrees with the one predicted by perturbation theory.

Again we can compute the first perturbative coefficients from the
previous Monte Carlo data. From the data on a $9^2$ lattice and
neglecting the contribution at four loops, we get
$p_4 = 6.09(3) \times 10^{-5}$. 
On $48^2$ we get $p_4 = 6.85(2) \times 10^{-5}$.

\vskip 1cm

\section{Conclusions}

\vskip 5mm

In this work we have checked the heating method to calculate
the renormalization constants present in the determination of
the topological susceptibility on the lattice. 
We simulated the $O(3)$ non-linear $\sigma$-model in two dimensions.
We used
the method at large correlation lengths in order to eliminate all
non-perturbative effects and see only the perturbation-theory
predictions on the Monte Carlo signal. The idea is based on the fact
that at fixed lattice size $L$ and large correlation lengths 
$\xi/L \rightarrow \infty$ perturbative calculations are expected to
become exact. This conjecture has been rigorously proven in the
$XY$ model \cite{lebowitz}. 
The positive conclusion of the test we performed supports
the conjecture also on the $O(3)$ model.

The choice of the $O(3)$ $\sigma$-model
was motivated by the fact that this
model can accommodate small instantons \cite{1ro}
and at very large correlation
length the size of the lattice in physical units strongly diminishes.

We computed the multiplicative renormalization $Z$ of the lattice
topological charge, $Q^{\rm latt}=QZ$ as well as the perturbative tail
of the corresponding topological susceptibility~$P$. Hence we
computed static quantities by using a Monte Carlo simulation thus
proving that the renormalization effects are the average of the
statistical fluctuations.

Concerning the calculation of the multiplicative renormalization $Z$
we know that one-instanton solutions do not exist on the
lattice and they are absent even for continuum tori \cite{rr}.
However we have checked that the usual solutions on the continuum
space-time \cite{poly} 
can be used to calculate $Z$ even though they present a
non-integer topological charge and are not stable solutions of the
lattice field equations. We think that this is a notable result.

Other conclusions which come out from our computation of $Z$ are
{\it (i)} the background charge $Q_0$ remains stable during the whole
heating trajectory, indeed we showed that the final result for $Z$ is
quite sensitive to the value of $Q_0$; {\it (ii)} we see that although
small fluctuations soon raise, small instantonic objects which could
lower the value of $Z$ are absent, thus the critical slowing down
seems to apply also to small instantons.

Another clear conclusion of this work is that to compute $Z$ we must
always divide the Monte Carlo signal $Q$ by the 
non-integer lattice 
topological charge $Q_0$ of the initial configuration.

A natural question raises at this point: does the $\rho$ dependence 
persist at values of $\beta$ in the scaling window ($\beta \sim
1.5 \div 2.5$)? At low values of $\beta$
there is more disorder in the configuration resulting in larger
statistical fluctuations which bring about larger error bars in the
lattice measure of any observable. On the other hand small topological
objects can appear \cite{inter} 
yielding a modification of the trajectories.
Therefore the answer of the previous question 
is possibly yes but the largest statistical
and systematic errors overwhelms this effect.

Finally, concerning the calculation of the perturbative tail, we saw a
good agreement between perturbation theory and Monte Carlo
results. They also displayed a lattice size dependence according to
the perturbation-theory predictions. 

The heating method works well and when correctly used it gives
the right answers. One has to take care also to the point where the
plateau sets in (in our simulations it sets in 
after many heating steps,
about 40 for the multiplicative renormalization)
and also to the 
strong autocorrelations which can mask the true plateau.

\vskip 1cm

\section{Acknowledgements}

\vskip 5mm

We thank Adriano Di Giacomo, Andrea Pelissetto and Ettore Vicari 
for useful
discussions. We also acknowledge financial support from the
Department of Physics of the University of Pisa and INFN.


\newpage

\noindent{\bf Figure captions}

\begin{enumerate}

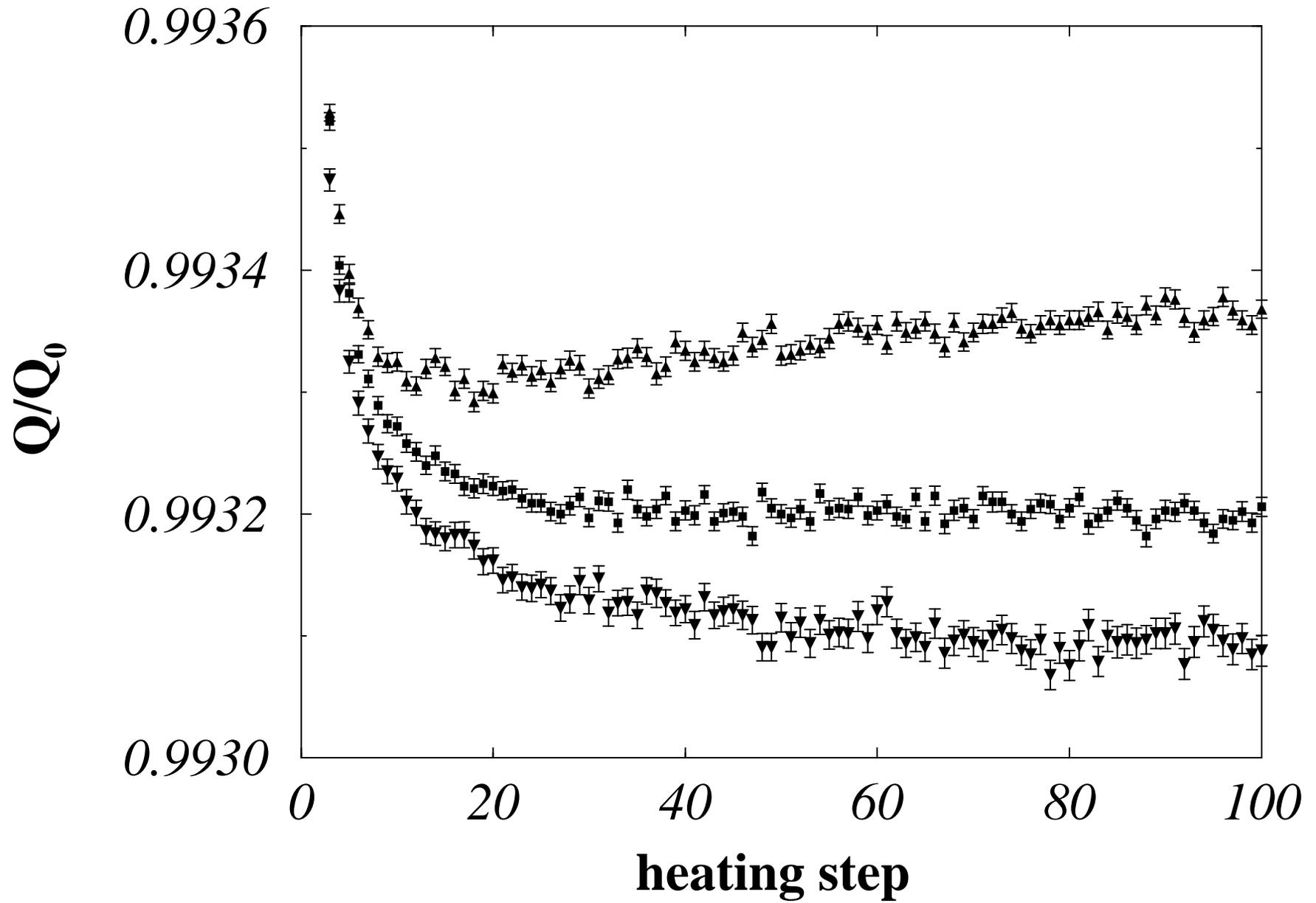
\item[Figure 1.] Monte Carlo signal for the topological charge after
100 heat-bath sweeps on a one
instanton background field. A lattice size $L=240a$ and $\beta=100$ was
used. The lower trajectory corresponds to $\rho=0.05L$ (down
triangles), the trajectory in the middle is for $\rho=0.13L$ 
(rectangles) and the upper curve is for $\rho=0.19L$ (up triangles).
A similar figure is obtained for the other lattice size used,
$L=1200a$.

\vskip 3mm

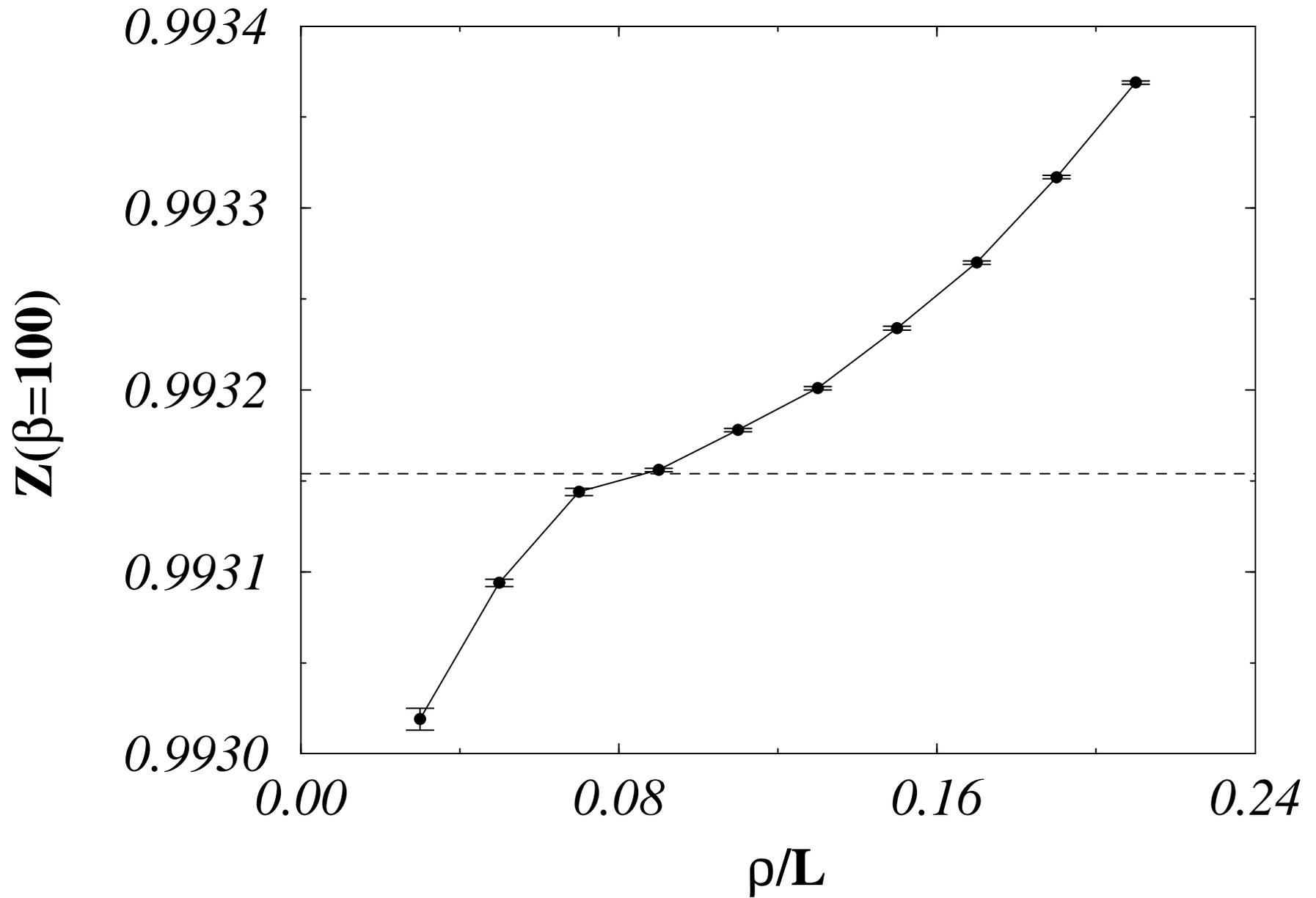
\item[Figure 2.] Values obtained for $Z$ on a $L=240a$ lattice and
$\beta=100$ as a function of the instanton size over $L$. The solid
line is to guide the eye. The dashed line is the 2-loop value.

\vskip 3mm

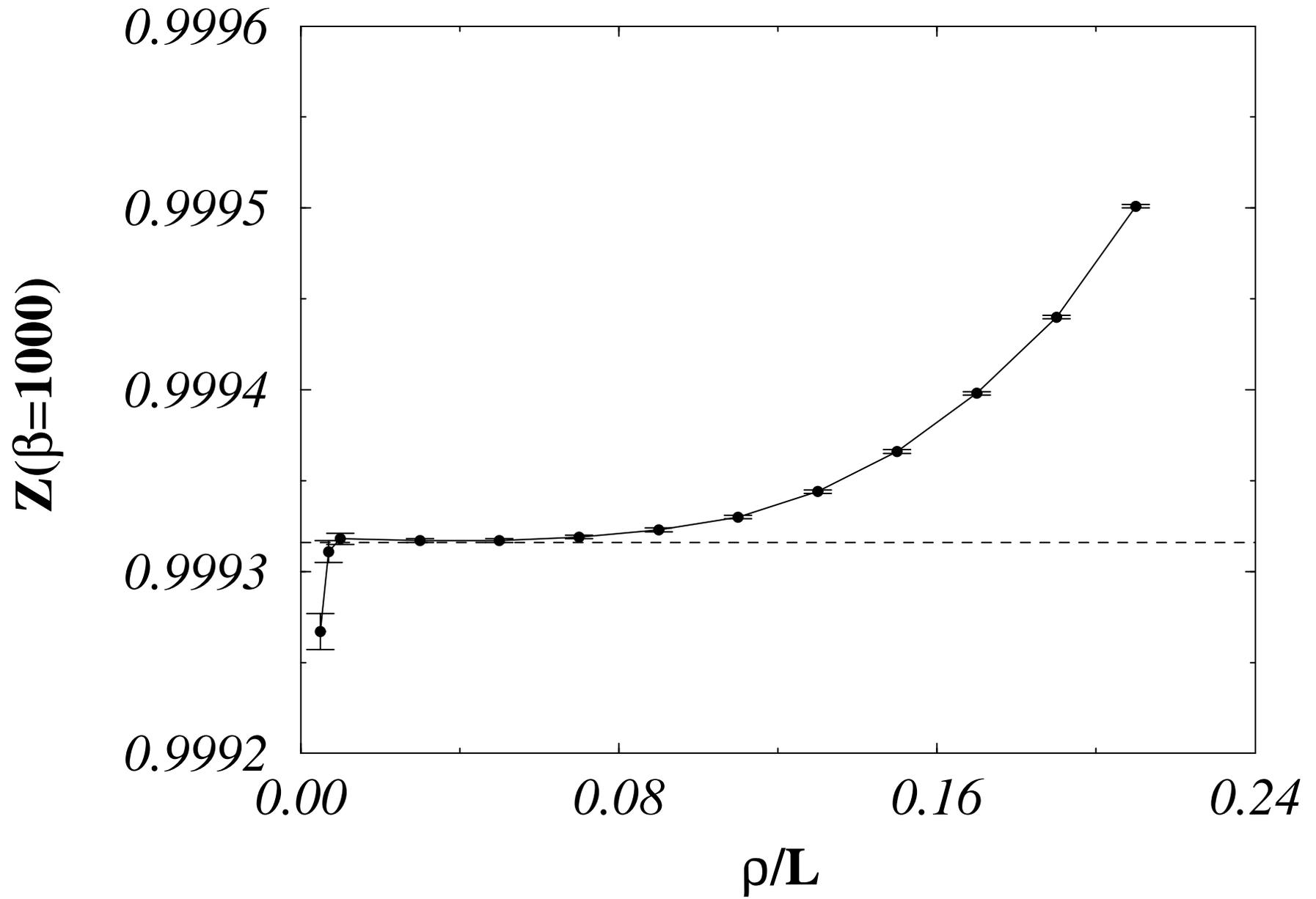
\item[Figure 3.] Values obtained for $Z$ on a $L=1200a$ lattice and
$\beta=1000$ as a function of the instanton size over $L$. The solid
line is to guide the eye. The dashed line is the 2-loop value.

\vskip 3mm

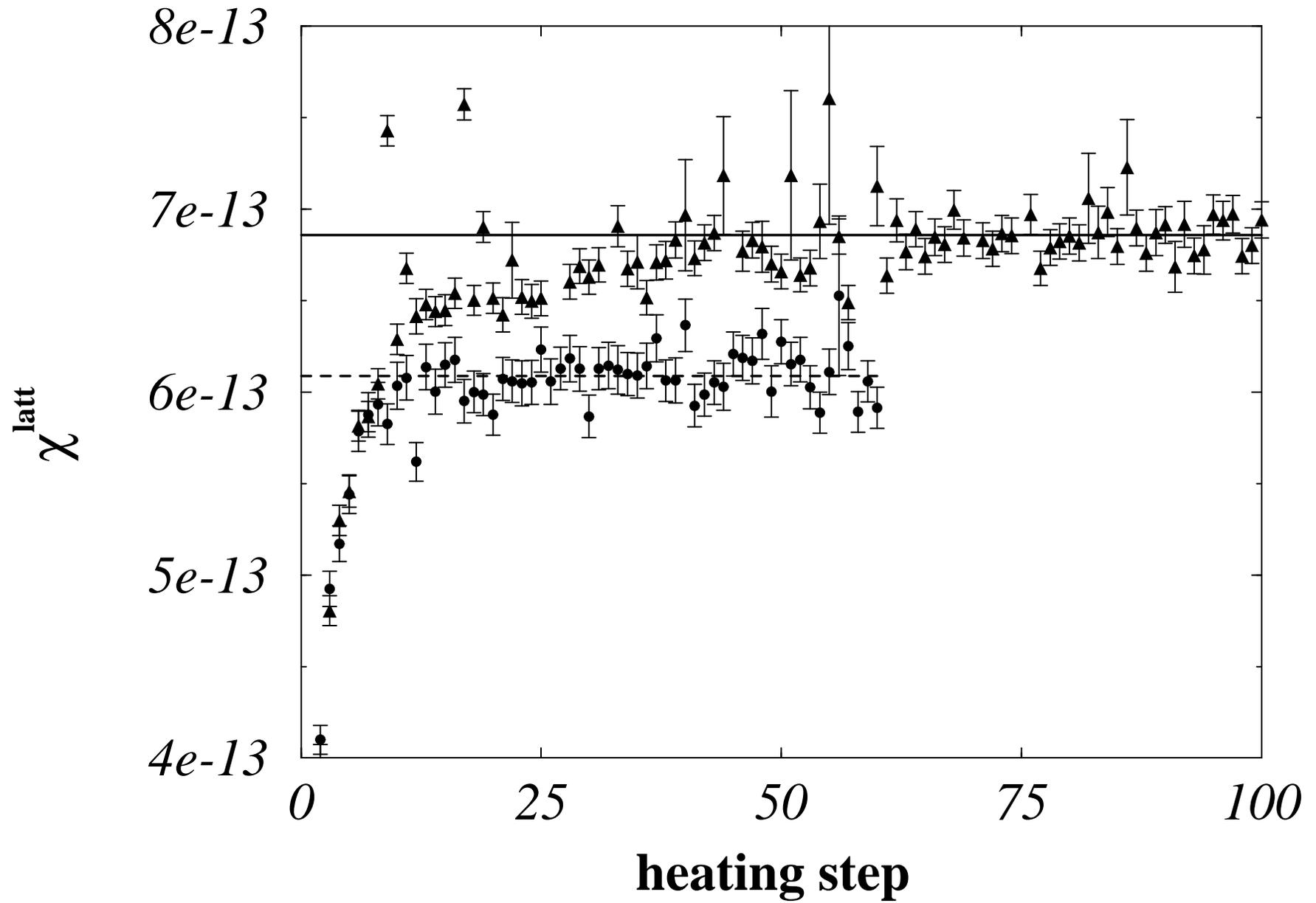
\item[Figure 4.] Monte Carlo signal of the topological susceptibility
at $\beta=100$. Triangles and solid line (circles and dashed line) are
the Monte Carlo signal and perturbative prediction on a $48^2$ 
($9^2$) lattice. On the $48^2$ ($9^2$) lattice, 100 (60)
heat-bath sweeps were performed.

\end{enumerate}

\vskip 2cm

\noindent{\bf Table captions}

\vskip 5mm

\begin{enumerate}

\item[Table I.] Values of $Z$ as a function of the instanton size
$\rho/L$. Results of a lattice size $L=240a$ and $\beta=100$.

\vskip 3mm

\item[Table II.] Values of $Z$ as a function of the instanton size
$\rho/L$. Results of a lattice size $L=1200a$ and $\beta=1000$.

\vskip 3mm

\item[Table III.] Flex point $\zeta$ 
obtained in the interpolation with a
polynomial of degree $p$. The interpolation was performed on the data
of Table I.

\end{enumerate}


\vskip 2cm

\centerline{\bf Table I}

\vskip 1cm

\moveright 2.4 in
\vbox{\offinterlineskip
\halign{\strut
\vrule \hfil\quad $#$ \hfil \quad & 
\vrule \hfil\quad $#$ \hfil \quad \vrule \cr
\noalign{\hrule}
\rho/L & Z(\beta=100) \cr
\noalign{\hrule}
0.03 & 0.993019(6) \cr
\noalign{\hrule}
0.05 & 0.993094(2) \cr
\noalign{\hrule}
0.07 & 0.993144(2) \cr
\noalign{\hrule}
0.09 & 0.993156(1) \cr
\noalign{\hrule}
0.11 & 0.993178(1) \cr
\noalign{\hrule}
0.13 & 0.993201(1) \cr
\noalign{\hrule}
0.15 & 0.993234(1) \cr
\noalign{\hrule}
0.17 & 0.993270(1) \cr
\noalign{\hrule}
0.19 & 0.993317(1) \cr
\noalign{\hrule}
0.21 & 0.993369(1) \cr
\noalign{\hrule}
}}

\vskip 2cm

\centerline{\bf Table II}

\vskip 1cm

\moveright 2.4 in
\vbox{\offinterlineskip
\halign{\strut
\vrule \hfil\quad $#$ \hfil \quad & 
\vrule \hfil\quad $#$ \hfil \quad \vrule \cr
\noalign{\hrule}
\rho/L & Z(\beta=1000) \cr
\noalign{\hrule}
0.005 & 0.999267(10) \cr
\noalign{\hrule}
0.007 & 0.999311(6) \cr
\noalign{\hrule}
0.01 & 0.999318(3) \cr
\noalign{\hrule}
0.03 & 0.999317(1) \cr
\noalign{\hrule}
0.05 & 0.999317(1) \cr
\noalign{\hrule}
0.07 & 0.999319(1) \cr
\noalign{\hrule}
0.09 & 0.999323(1) \cr
\noalign{\hrule}
0.11 & 0.999330(1) \cr
\noalign{\hrule}
0.13 & 0.999344(1) \cr
\noalign{\hrule}
0.15 & 0.999366(1) \cr
\noalign{\hrule}
0.17 & 0.999398(1) \cr
\noalign{\hrule}
0.19 & 0.999440(1) \cr
\noalign{\hrule}
0.21 & 0.999501(1) \cr
\noalign{\hrule}
}}

\vskip 2cm

\centerline{\bf Table III}

\vskip 1cm

\moveright 2.4 in
\vbox{\offinterlineskip
\halign{\strut
\vrule \hfil\quad $#$ \hfil \quad & 
\vrule \hfil\quad $#$ \hfil \quad \vrule \cr
\noalign{\hrule}
 p & \zeta \cr
\noalign{\hrule}
3 & 0.993197 \cr
\noalign{\hrule}
5 & 0.993162 \cr
\noalign{\hrule}
7 & 0.993154 \cr
\noalign{\hrule}
9 & 0.993159 \cr
\noalign{\hrule}
}}

\end{document}